\documentclass[final,5p,times,twocolumn]{elsarticle}

\usepackage[numbers]{natbib}

\usepackage{amsmath,amssymb,amsfonts}
\usepackage{graphicx}
\usepackage{psfrag}
\usepackage{multicol}
\usepackage{multirow}
\usepackage{booktabs}
\usepackage{xcolor}
\usepackage[utf8]{inputenc}

\usepackage{cleveref}
\crefname{section}{§}{§§}
\Crefname{section}{§}{§§}

\usepackage[switch]{lineno}

\begin{document}

\begin{frontmatter}
\title{\LARGE \bf
Model Predictive Control (MPC) of an Artificial Pancreas with Data-Driven Learning of Multi-Step-Ahead Blood Glucose Predictors
}

\author[unitn]{Eleonora Maria Aiello }
\author[hu]{Mehrad Jaloli}
\author[hu]{Marzia Cescon\corref{cor1} }

\address[unitn]{Department of Computer Science and Engineering, University of Trento, Povo di Trento, 38123, Italy.}
\address[hu]{Department of Mechanical Engineering, University of Houston, Houston, TX 77004, USA}

\cortext[cor1]{Corresponding author Marzia Cescon\\ E-mail address: mcescon2@uh.edu}

%
\begin{abstract}
We present the design and \textit{in-silico} evaluation of a closed-loop insulin delivery algorithm to treat type 1 diabetes (T1D) consisting in a data-driven multi-step-ahead blood glucose (BG) predictor integrated into a Linear Time-Varying (LTV) Model Predictive Control (MPC) framework. Instead of identifying an open-loop model of the glucoregulatory system from available data, we propose to directly fit the entire BG prediction over a predefined prediction horizon to be used in the MPC, as a nonlinear function of past input-ouput data and an affine function of future insulin control inputs. For the nonlinear part, a Long Short-Term Memory (LSTM) network is proposed, while for the affine component a linear regression model is chosen.  To assess benefits and drawbacks when compared to a traditional linear MPC based on an auto-regressive with exogenous (ARX) input model identified from data, we evaluated the proposed LSTM-MPC controller in three simulation scenarios: a nominal case with 3 meals per day, a random meal disturbances case where meals were generated with a recently published meal generator, and a case with 25$\%$ decrease in the insulin sensitivity. Further, in all the scenarios, no feedforward meal bolus was administered. For the more challenging random meal generation scenario, the mean $\pm$ standard deviation percent time in the range 70-180 [mg/dL] was 74.99 $\pm$ 7.09 vs. 54.15 $\pm$ 14.89, the mean $\pm$ standard deviation percent time in the tighter range 70-140 [mg/dL] was 47.78$\pm$8.55 vs.  34.62 $\pm$9.04, while the mean $\pm$ standard deviation percent time in sever hypoglycemia, i.e., $<$ 54 [mg/dl] was 1.00$\pm$3.18 vs. 9.45$\pm$11.71, for our proposed LSTM-MPC controller and the traditional ARX-MPC, respectively. Our approach provided accurate predictions of future glucose concentrations and good closed-loop performances of the overall MPC controller.

\vspace{5pt}
\end{abstract} 
\end{frontmatter}

\section{INTRODUCTION}
Type 1 diabetes (T1D) is a metabolic condition characterized by high blood glucose levels (hyperglycemia), caused by the autoimmune irreversible destruction of the pancreatic $\beta$-cells, which are responsible for the production and release of the hormone insulin. The chronic diabetes hyperglycemia, i.e. blood glucose (BG) levels $\geq$ 180 mg/dL, leads to an increased risk of life-threatening events, such as diabetes ketoacidosis, and has serious long-term complications associated with damage, dysfunction and failure of various organs~\cite{2014Daco}. Exogenous insulin is therefore required for individuals with T1D to adequately regulate their BG concentration in the euglycemic range~\cite{2018PAtG}, i.e. 70-180 mg/dL. Despite burdensome insulin treatment, however, individuals with T1D experience difficulties in maintaining healthy BG levels and fail to meet the recommended glycemic targets. Over the past 40 years, significant effort has been directed toward the automated control of blood glucose concentration, and thanks to the recent technological advances in glucose sensing devices and insulin infusion mechanisms, effective glucose regulation is becoming increasingly possible~\cite{CSM2018,DoyleFrancisJ2014Caps,ThabitHood2016Coat,HaidarAhmad2016TAPH, Lal2019, Aiello2021}. Model predictive control (MPC)~\cite{maciejowski2002predictive,rawlings2017model} is an attractive control strategy for closed-loop insulin delivery and has been considered in this context by various authors showing promising results in the management of diabetes in a hybrid fashion (so-called hybrid closed-loop systems)~\cite{del2014first,buckingham2018safety,brown2019six,brown2021multicenter,tauschmann2016home,garg2017glucose,deshpande2022feasibility,ozaslan2022feasibility}. In the hybrid setting, the MPC algorithm is in charge of adjusting the clinical defined  basal insulin profile during fasting periods based on real-time measurements from continuous glucose monitoring (CGM) device, while a linear feed-forward control action is included in the control scheme based on the announcement of the disturbances provided by the user. 
The MPC algorithms currently adopted in clinical trials rely on linear models to describe the process under control, to optimize the control performance and ensure constraint satisfaction over a prediction horizon \cite{rawlings2017model}. 
In the authors' opinion, from the algorithmic point of view, the major limitation affecting glucose control schemes, lies in the inaccuracies of the linear models used for generating the BG predictions. The use of a linear model is justified by its simplicity and small computational load, but a linear model can only approximate the complex non-linear dynamic of the human metabolism. 
Specifically, based on the author's experience, long prediction horizons of the same length of hyperglycemic perturbations, e.g., after a mixed meal, introduce large errors due to large plant-model mismatch, and are therefore not advantageous. However, on the other hand, shorter prediction horizons yield control laws that persist to command insulin delivery beyond what is required, especially when a hyperglycemic state is in the process of correcting itself. Similar observations can be made when dealing with glycemic perturbations arising from stress, illness, hormonal variations and physical activity, which are not captured by the available linear models in the literature. The resulting insulin administration is inadequate and contributes to the risk of dangerous controller-induced hypoglycemia and increased time outside of the target range.
That said, in this contribution we explore a different way of approaching the modeling step in a linear time-varying (LTV) MPC framework. In particular, we consider the estimation of long-term horizons multi-step ahead predictors of BG dynamics for receding horizon control starting from input-output data, with the goal of increasing the prediction accuracy for longer prediction horizons. Given the inherent physiological nonlinearity in the underlying glucose metabolism, data-driven approaches based on Artificial Neural Networks (ANNs) can achieve remarkable performance over traditional linear models, as shown in our previous works~\cite{AIELLO2020103255,jaloli2022}, thanks to their ability to perform automatic feature extraction, and hence eliminating the need of feature engineering \cite{zhu2018deep, li2019glunet}. 
In particular, unlike traditional Recursive Neural Networks (RNNs), Long Short Term Memory (LSTM) networks are specifically designed to learn and retain information over long sequences~\cite{greff2016lstm, gers2000learning} and were successfully applied for blood glucose predictions \cite{AIELLO2020103255,jaloli2022,mirshekarian2017using, carrillo2021long,  iacono2022personalized}.
In this work we build our predictors as the superposition of a nonlinear function of past inputs and past outputs and an affine function of future control moves. This predictor construction, first proposed in \cite{masti2020learning}, allows us to solve the MPC problem via quadratic programming (QP) despite the nonlinearity of the system. We propose a LSTM network for the nonlinear part, and a linear regression model for the affine component of the predictor.

The performance of the proposed approach is demonstrated on three real-life use-case scenarios using a widely accepted metabolic model of glucose metabolism equipped with ten in-silico adult subjects~\cite{kovatchev2020method} and compared against a linear autoregressive exogenous (ARX) model. 

The remainder of the paper is structured as follows. In Sec.~\ref{sec:predictors} we introduce the multi-step ahead BG predictors; the proposed MPC control law is outlined in Sec.~\ref{sec:mpc}; while in Sec.~\ref{sec:evaluation} we describe the simulation scenarios used to evaluate control performances. Section~\ref{sec:results} shows results and finally Sec.~\ref{sec:summary} concludes the paper.



\section{DATA-DRIVEN MULTI-STEP-AHEAD BLOOD GLUCOSE PREDICTORS FOR MPC}\label{sec:predictors}

\subsection{A predictor structure affine in the future inputs}
Let an unknown, non-linear discrete-time system $\Sigma$ be described by the dynamical model:
\begin{equation}
\Sigma = 
\begin{cases}
x_{k+1} = f_\Sigma(x_k,u_k)\\
y_k = h_\Sigma(x_k)
\end{cases}
\end{equation} 
where $k$ is the discrete-time instant, $x_k \in \mathbb{R}^{n_x}$ the state of the system at time $k$, $u_k \in \mathbb{R}^{n_u}$ and $y_k \in \mathbb{R}^{n_y}$ input and output  at time $k$, respectively, $f_\Sigma: \mathbb{R}^{n_u \times n_x}\rightarrow \mathbb{R}^{n_x}$, and $h_\Sigma: \mathbb{R}^{n_x}\rightarrow \mathbb{R}^{n_y}$. 

Our aim is to design $T$ predictors of the system output $y_{k+j}$, based on input-output data available from the system $\Sigma$ at time $k$, where $j = 1,\dots,T$ denotes the prediction step and $T \in \mathbb{N}$ is the prediction horizon in the MPC problem. Note that here we propose to learn one predictor per each prediction step $j$, as opposed to the traditional approach in the process control literature prescribing the iteration of the system equations for the derivation of output predictions beyond one-step ahead. Stacking all the predicted sequences on top of each other, we obtain:
\begin{equation}\label{eq:predictor}
     \hat{Y}_T = H_T(x_k, U_{T-1}).
\end{equation}
where $H_T : \mathbb{R}^{{n_x} \times {Tn_u}}  \rightarrow \mathbb{R}^{Tn_y}$ is the multi-step predictor, comprised of one predictor for each prediction step, $\hat{Y}_T=[\hat{y}_{k+1}, \dots, \hat{y}_{k+T}]'$, is the predicted output sequence, and $U_{T-1}=[u_{k}, \dots, u_{k+T-1}]'$ is the future input sequence, along the prediction horizon.

The multi-step predictor described in Eq.~\ref{eq:predictor} is intended to be used as the process model for a MPC formulation. Note that, when choosing a quadratic cost function and linear constraints in the MPC, an input-affine formulation of the predictor is preferable, as the resulting Finite Horizon Optimal Control Problem (FHOCP) in this case turns out to be a quadratic program, which can be efficiently solved using ad-hoc solvers \cite{rawlings2017model,masti2020learning}. For this reason, the general multi-step predictor formulation of Eq.\ref{eq:predictor} is then specialized to an affine form with respect to the future control moves  \cite{masti2020learning}:
\begin{equation}
\label{eq:tt}
\hat{Y}_T = F_T(x_k) + G_T(x_k)({U}_{T-1}-\bar{U}_{T-1})
\end{equation}
where $\bar{U}_{T-1}$ is a nominal input sequence and is assumed to be equal to the steady-state nominal input, $F_T:\mathbb{R}^{{n_x} }  \rightarrow \mathbb{R}^{Tn_y}$  and $G_T :\mathbb{R}^{{n_x} }  \rightarrow \mathbb{R}^{{Tn_y}\times T{n_u}}$.
Additionally, to reduce the predictor complexity, a recursive structure is adopted, with $F_T$ and $G_T$ defined as follows  \cite{masti2020learning}:
 \begin{equation} \label{eq:affinePredictor}
 F_T=\left[
 \begin{array}{c}
      f_1(x_k)  \\
      f_2(x_k)   \\
      \vdots \\
      f_{T-1}(x_k) \\
      f_{T}(x_k) 
 \end{array}
 \right], 
\end{equation}
 \begin{equation} \label{eq:affinePredictorG}
 G_{T}=\left[
 \begin{array}{ccccc}
      g_1(x_k) & 0 & \cdots & 0 & 0 \\
      g_1(x_k) & g_2(x_k) & \cdots & 0 & 0  \\
      \vdots \\
     g_1(x_k) & g_2(x_k) & \cdots & g_{T-1}(x_k) & 0 \\
     g_1(x_k) & g_2(x_k) & \cdots & g_{T-1}(x_k) & g_{T}(x_k)
 \end{array}
\right]
\end{equation}
In this work, the state vector $x_k$ contains past and current CGM data ($y^{cgm}$), the past delivered insulin ($u^{ins}$) and the past and current ingested carbohydrates ($d^{cho}$):

\begin{align}
\label{eq:xk}
    x_k=&[y^{cgm}_k, y^{cgm}_{k-1}, \dots, y^{cgm}_{k-3*T} \\
    & u^{ins}_{k-1}, u^{ins}_{k-2}, \dots, u^{ins}_{k-3*T-1} \\
    & d^{cho}_{k}, d^{cho}_{k-1}, \dots, d^{cho}_{k-3*T} ]
\end{align} 

The nominal input sequence $\bar{U}_{T-1}$ is chosen to be the basal insulin, which is a  constant amount of insulin in charge of maintaining the glucose levels in the euglycemic range during fasting periods. Insulin variations, ${U}_{T-1}-\bar{U}_{T-1}$, are obtained by injecting insulin boluses, and aim at minimizing the occurrence of hypo- and hyperglycemia in presence of disturbances (e.g., meals, physical activity). 

\subsection{Learning the predictors}

The problem of learning the maps $F_T$ and $G_T$ in Eqs.~\ref{eq:affinePredictor}-\ref{eq:affinePredictorG} can be posed as two separate learning problems, and the prediction $\hat{Y}_T$ can be seen as the sum of two contributions, which we propose to identify separately.   
In order to ease the two-steps identification procedure, we generated two training datasets \textit{in-silico} using the metabolic model and the 10-subject adult cohort in \cite{kovatchev2020method} with different insulin therapies:
\begin{itemize}
    \item \textit{Scenario-I}: Insulin therapy based only on the administration of the basal insulin throughout the day.
    \item \textit{Scenario-II}: Insulin therapy based on the administration of basal insulin together with insulin boluses at mealtimes.
\end{itemize} 
Both scenarios have the same sequence of meals, generated with a recently published meal generator~\cite{aiello2022model}, however only \textit{Scenario-II} includes an insulin bolus at mealtime to compensate for the induced prandial glucose rise. Following conventional therapy, the insulin bolus dose is computed as the ratio between the meal amount and the carbohydrate-to-insulin (CR) factor~\cite{walsh2006pumping}. 
Since \textit{Scenario-I} corresponds to the condition of having ${U}_{T-1}=\bar{U}_{T-1}$, we used this dataset to identify $F_T$, which models the dynamics of the system from past input-output sequences collected in quasi open-loop conditions, i.e., without applying any control action to reject meal disturbances.
On the other side, the dataset generated with the second insulin therapy was exploited to identify $G_T$, which represents the effect of any control action on the glucose predictions.
The sampling time $T_s$ chosen for the overall $T$-step-ahead predictor is 15 minutes, which provides a sufficiently accurate reproduction of the continuous time dynamics of the system and allows spanning over a long time with a reasonable number of predictors. The system under control typically settles in 2 to 4 hours \cite{american2001postprandial}. 

\subsubsection{Modelling $F_T$ via LSTMs}

\begin{figure}[t]
        \centering
        \includegraphics[width=0.48\textwidth]{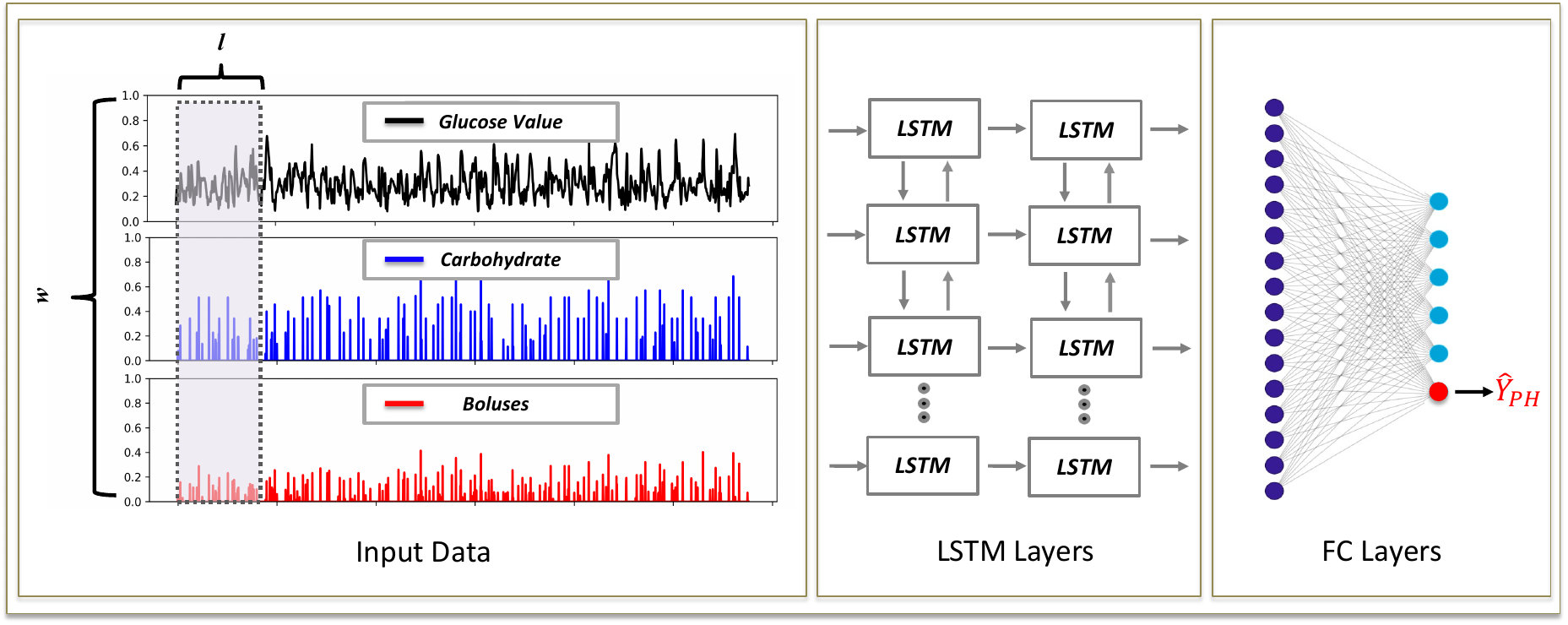}
        \caption{Proposed Model Architecture for Blood Glucose Prediction. Input samples are processed through LSTM layers to capture temporal dependencies, and the output is mapped to a suitable format for prediction using fully connected (FC) layers}
    \label{fig:LSTM}
\end{figure} 
We designed the $T$ functions $f_j$, with $j=1,\dots, T$, as a 2-layer stacked LSTM model as illustrated in Fig.~\ref{fig:LSTM}. As described in the  previous section, each predictor is trained using the synthetic dataset where the insulin therapy includes only the administration of the basal insulin, i.e. \textit{Scenario-I}. This assumption implies that $U_{T-1}=\bar{U}_{T-1}$, and consequently, $\hat{Y}_T=F({x_k})$. 


Each LSTM layer receives input from the previous layer, enabling the model to learn hierarchical representations of the data. 
The output of the last LSTM layer is then passed to a stack of two fully connected (FC) layers. The FC layers are responsible for mapping the LSTM's hidden representations to a suitable format for prediction. They introduce additional non-linearity and complexity to the model, enabling it to capture more intricate patterns in the data. 
The output size of each predictor $f_j$ is set equal to $j$ such that the number of neurons in the output layer equals the number of samples up to $j$. Finally, as we combine each $f_j$ into a $T$-step-ahead predictor, then $\hat{y}_{k+j}$ is the j$^{th}$ sample from $f_j(x_k)$.
The training process uses a Mean Absolute Error (MAE) loss function, with the Root Mean Square Propagation (RMSProp) algorithm to minimize the loss function and update the model's parameters. The MAE loss function is defined as: 
\begin{equation}
L_{MAE}=\frac{\sum_{i=1}^{n}|y_i- \hat{y}_i |} {n}
\end{equation}
where $L_{MAE}$ is the loss, $n$ represents the number of samples in the dataset,  $y_i$ and $\hat{y}_i$ are the actual and the predicted glucose samples, respectively. We set the learning rate to 0.0001 to ensure stable and effective convergence during training.

The forward chaining methodology was applied for training and validation of each $f_j$. This approach is specifically designed to handle temporal dependencies and ensures realistic evaluation of the model's performance \cite{bergmeir2012use}.
Forward chaining involves iteratively training the model on a subset of the training data and testing it on unseen future data. In this work, at each iteration $k_p$, with $k_p=1,\dots,9$, the model is trained on data from subject 1 to subject $k_p$ and is then validated on subject $k_p+1$, representing unseen future data. This validation step assesses the model's ability to generalize and make accurate predictions for new patients. 
To enhance the performance of our LSTM model for blood glucose prediction, we undertook a comprehensive process involving hyperparameter tuning, early stopping, and meticulous model evaluation.  In the pursuit of optimal hyperparameters, we explored different batch sizes (64, 128, and 256) and the number of training epochs (ranging from 200 to 400). These selections were made based on the datasets' characteristics, considering the length of the dataset and the number of patients involved in training \cite{hastie2009elements}. This exploration allowed us to identify the most effective combination of batch size and training epochs that yielded superior predictive capabilities. To mitigate the risk of overfitting and promote generalization, we implemented the early-stopping-point approach. 
%
%
%
\subsubsection{Modelling $G_T$ via a linear regression model}
We modelled the $T$ functions $g_j$, with $j=1,\dots, T$, as linear regression models able to describe the relation between glucose concentration and injected bolus insulin.
As described in Section 2.1, each component $g_j$ has to model the effect of any control action on the glucose predictions. Specifically, $G_T$ has to describe the glucose response to the variation in insulin dose with respect to the basal insulin rate, defined as $\Delta U := U_{T-1}-\bar{U}_{T-1}$. In order to capture this relationship, we rely on the concept of the correction factor (CF), inspired by the work proposed in \cite{aiello2019improving}. 
The CF is a clinical parameter that describes how much 1 unit of rapid-acting insulin will reduce the blood glucose from the current level. The CF is an estimate of the so-called insulin sensitivity which is unknown, but it affects the observed blood glucose trends \cite{davidson2008analysis}. That is the following is assumed: the flatter the glucose trend, the lower the insulin sensitivity. 
Assuming $F_T$ has been identified and hence available, the mismatch between the glucose data generated in \textit{Scenario-II} and the predicted open-loop glucose excursions can be calculated as:
\begin{equation}
  \Delta Y^{cgm,F_T}= Y^{cgm}-F_T(x_k)
\end{equation} 
where $\Delta Y^{cgm,F_T}=[\Delta y^{cgm}_{k+1}, \Delta y^{cgm}_{k+2}, \dots, \Delta y^{cgm}_{k+T}]'$ 
are the error samples at time instant $k$, and $Y^{cgm}=[y^{cgm}_{k+1}, y^{cgm}_{k+2}, \dots, y^{cgm}_{k+T}]'$ are the actual CGM samples at time instant $k$. 
From Eq.~\ref{eq:tt}, we can derive that $G_T$ describes the ratio between $\Delta Y^{cgm,F_T}$ and $\Delta U$. This implies that $G_T$ plays the role of the correction factor. Since the correction factor can be indirectly estimated by observing the past glucose trend, the proposed regressors for $G_T$ are the past and current CGM data, which are a subset of the variables of the state vector $x_k$, as described in Eq.~\ref{eq:xk}:
\begin{equation}
    \Delta Y^{cgm,F_T}=G_T \cdot \Delta U   
\end{equation}
Because of the recursive structure of $G_T$ described in Eq. \ref{eq:affinePredictorG}, we obtain the following structure:
\begin{equation}   
G_T= 
\left[
 \begin{array}{cccc}
      C_1  & 0 & \cdots & 0 \\
      C_1  & C_2& \cdots & 0    \\
      \vdots \\
       C_1 & C_2 &  \cdots & C_T
 \end{array}
 \right] 
 \left[
  \begin{array}{cccc}
    Y^{cgm}_k   & 0 & \cdots & 0 \\
      Y^{cgm}_k  & Y^{cgm}_k   & \cdots & 0    \\
      \vdots \\
       Y^{cgm}_k  & Y^{cgm}_k&  \cdots & Y^{cgm}_k
 \end{array}
 \right]
\end{equation}
where $C_j=[c_{j,0}, \dots, c_{j,3T}]$ with $j=1,\dots,T$, $Y^{cgm}_k=[y^{cgm}_{k}, \dots, y^{cgm}_{k-3\cdot T} ]$. To identify $G_T$, we used the synthetic dataset from \textit{Scenario-II}, where the insulin therapy includes both the administration of basal insulin and the insulin boluses at mealtimes.  

\section{IMPLEMENTATION IN A MPC}\label{sec:mpc}

Let $J(\cdot)$ be the cost function of the MPC problem:
%
\begin{align}
J(x_k,U_{T-1},&\bar{Y}_T, \bar{U}_{T-1})=(\hat{Y}_{T}(x_k)-\bar{Y}_T)'{Q}(\hat{Y}_{T}(x_k)-\bar{Y}_T)\\
&+(U'_{T-1}-\bar{U}_{T-1}){R}(U_{T-1}-\bar{U}_{T-1})
\end{align}
\label{eq:J}
where $T>0$ is the prediction horizon, $\bar{Y}_T$ the glucose set point prediction along the horizon $T$, $Q=qI_T$ and $R=rI_T$ are tuning parameters, $q>0$, $r>0$, $I_T\in \mathbb{R}^{T \times T}$ is the identity matrix. The basal insulin $\bar{U}_{T-1}$ is the subject-dependent basal insulin rate. 
A time-dependent set point $\bar{Y}_T$ was employed and set to 110 mg/dl during the day, and 125 mg/dl at night. Daytime is defined to be the interval 5:00 am to 10:00 pm. All other times are nighttime and there is no transition period \cite{gondhalekar2013periodic}. This trade-off  helps  to  reduce  the  chance  of  immediately  dangerous nocturnal hypoglycemic events, i.e. glucose levels below 70 mg/dL. The proposed MPC law uses T := 8, i.e. 120 minutes, q:= $1$, r:= $10$. Then, the resulting FHOCP is reported in (\ref{eq:MPC})-(\ref{eq:constr}):
\begin{equation}
U^*_{T-1}(x_k, \bar{Y}_T, \bar{U}_{T-1} )=\text{arg} \min_{U_{T-1}}\;J(x_k,U_{T-1},\bar{Y}_T, \bar{U}_{T-1} )
\label{eq:MPC}
\end{equation}
\textit{subject to:}
\begin{equation}
\label{eq:dynLin}
\begin{array}{c}
\hat{Y}_{T}(x_k)= F_T(x_k) + G_T(x_k)({U}_{T-1}-\bar{U}_{T-1})\\
\end{array}
\end{equation}
\begin{equation}
\label{eq:constr}
\begin{array}{c}
Y_{min}  \leq  \hat{Y}_{T}(x_k) \leq Y_{max}\\
U_{min}\leq U_{T-1} \leq U_{max}\\
\end{array}
\end{equation}
The constraint values are defined according to hardware limitations of the insulin pump and the CGM:

\begin{equation}
\begin{array}{l}
Y_{min}=0 \;mg/dl\;\;\;Y_{max}=500\;mg/dl\\
U_{min}=0 \;U\;\;\;\;\;\;\;\;U_{max}=25\;U\\
\end{array}
\label{eq:constraints}
\end{equation}

\noindent The receding horizon (RH) principle is then applied so that the actual control action given as output, $u_{mpc}$, is the first element of the optimal control sequence $U^{*}_{T-1}$ at each time instant $k$.

\section{IN-SILICO PERFORMANCE EVALUATION}\label{sec:evaluation}
\subsection{Evaluation of the predictors}
In this paper, two different scenarios were designed for the identification, i.e., \textit{Scenario-I} and \textit{Scenario-II}, and a separate scenario (\textit{Scenario-III}) was designed for the validation of the $T$-step-ahead predictors. 
In both identification and validation scenarios, the mealtimes and corresponding amounts were defined by using a stochastic meal generator, which was designed for an accurate reproducibility of eating habits of a cohort of subjects with T1D \cite{aiello2022model}. The stochastic meal generator is based on a Markov Chain, whose state is the fasting period and the transition probabilities depend on daytime and carbohydrate intake of the previous meal. The use of a stochastic generator choice removes the arbitrariness of the scenario design process and provides enough variability in the synthetic data. As a consequence, identification and validation scenarios have different sequences of meals: different food habits imply different insulin therapies, which in turn impact differently on glucose levels, as observed in standard clinical practice. 
All scenarios last 28 days.  

In this work, we considered a prediction horizon 
$T=120$ minutes, with a sampling time of $T_s=15$ minutes. Denoting with
$\hat{y}$ the model prediction and
$y$ the measured glucose level in the validation scenario, the accuracy of the model predictions of each single predictor composing the overall $T$-step-ahead predictor, is computed in terms of mean absolute error (MAE), mean absolute percentage error (MAPE) and root mean squared error (RMSE), defined as:
\begin{equation}
\text{MAE} = \frac{1}{n} \sum_{i=1}^{n} |y_i - \hat{y}_{i}|
\end{equation}
\begin{equation}
\text{MAPE} = \frac{1}{n} \sum_{i=1}^{n} \left|\frac{y_i - \hat{y}_{i}}{y_i}\right| \times 100\%
\end{equation}
\begin{equation}
\text{RMSE} = \sqrt{\frac{1}{n} \sum_{i=1}^{n} (y_i - \hat{y}_{i})^2}.
\end{equation} 
\begin{table*}[h!]
\caption{Accuracy metrics, including mean absolute error (MAE), mean absolute percentage error (MAPE) and root mean squared
error (RMSE), of the proposed multi-step predictor vs. that of the ARX model-based predictor. Data is shown as mean (standard deviation) over the \textit{in-silico} population on validation dataset.}
  \label{tab:prediction_performances}
  \centering
  \begin{tabular}{c|c|c c|c c|c c}
     \multirow{2}{*}{$j $ } &
    \multirow{2}{*}{$T_s \cdot j $ [min]} &
      \multicolumn{2}{c}{MAE [mg/dL]} &
      \multicolumn{2}{c}{MAPE [$\%$]} &
      \multicolumn{2}{c}{RMSE [mg/dL]} \\
   & & Multi-step & ARX & Multi-step & ARX & Multi-step  & ARX \\
    \hline \hline
    1&15 & 1.52(0.27) & 1.75(0.20) & 0.85(0.11) & 1.04(0.06) & 2.18(0.44) & 2.11(0.25)\\
    
    2&30 & 3.97(0.75) & 5.33(0.63) & 2.18(0.40) & 3.15(0.17) & 5.33(1.13) & 6.24(0.73) \\
    
    3&45 & 7.82(1.81) & 10.21(1.19) & 4.26(1.04) & 6.00(0.27) & 11.44(2.78) & 11.73(1.44) \\
    
    4&60 & 14.14(2.86) & 15.72(1.86) & 7.62(1.78) & 9.21(0.33) & 22.33(4.83) & 17.91(2.32) \\
    
    5&75 & 20.98(4.30) & 21.42(2.69) & 11.04(2.71) & 12.52(0.42) & 33.82(7.52) & 24.21(3.30) \\
    
    6& 90 & 27.07(5.75) & 26.97(3.60) & 14.07(3.63) & 15.72(0.56) & 41.89(9.68) & 30.25(4.25) \\
    
    7& 105 & 28.68(6.22) & 32.18(4.51) & 15.05(3.97) & 18.73(0.75) & 39.38(8.56) & 35.86(5.13) \\
    
    8& 120 & 35.32(7.22) & 37.01(5.32) & 18.32(4.52) & 21.50(0.92) & 50.91(11.26) & 40.95(5.92)\\
    \bottomrule
  \end{tabular}
\end{table*}
\subsection{Evaluation of the controller}
The evaluation of control performances was carried out using the 10 \textit{in-silico} subjects. Simulations were 48 h in duration, starting at midnight.  With the aim to test our controller in different but realistic conditions, including real-life food habits and unknown changes in insulin sensitivity, we considered three scenarios:
\begin{itemize}
	\item \textit{Scenario A}: a comparatively standard scenario where each subject consumes three meals per day, including breakfast of 50 g at 8 AM, lunch of 75 g at 1 PM, dinner of 75 g at 7 PM \cite{gondhalekar2018velocity}
	\item \textit{Scenario B}: this scenario aims at mimicking realistic meal habits, with meal time and amounts defined by using the stochastic meal generator proposed in\cite{aiello2022model} 
	\item \textit{Scenario C}: in this scenario, additional insulin resistance is introduced by 25\% of the nominal values. The controller is unaware of this change in the insulin sensitivity.     
	\end{itemize}

We evaluated the controller for each scenario based on the metrics that are considered most relevant in clinical practice \cite{battelino2019clinical}, including mean glucose (mg/dL), glycemic variability, i.e. coefficient of variation (CV), (mg/dl), percent time above 250 mg/dL, percent time above 180 mg/dL, percent time in range, i.e. between 70–180 mg/dL, percent time below 70 mg/dl, and percent time below 54 mg/dl. Recommended glucose targets include percent time in range of 70-180 mg/dL $>70\%$, percent time below 70 mg/dL $<4\%$, and percent time above 180 mg/dL $<25\%$ \cite{battelino2019clinical}.
\begin{figure*}
        \centering
    \includegraphics[width=0.48\textwidth]{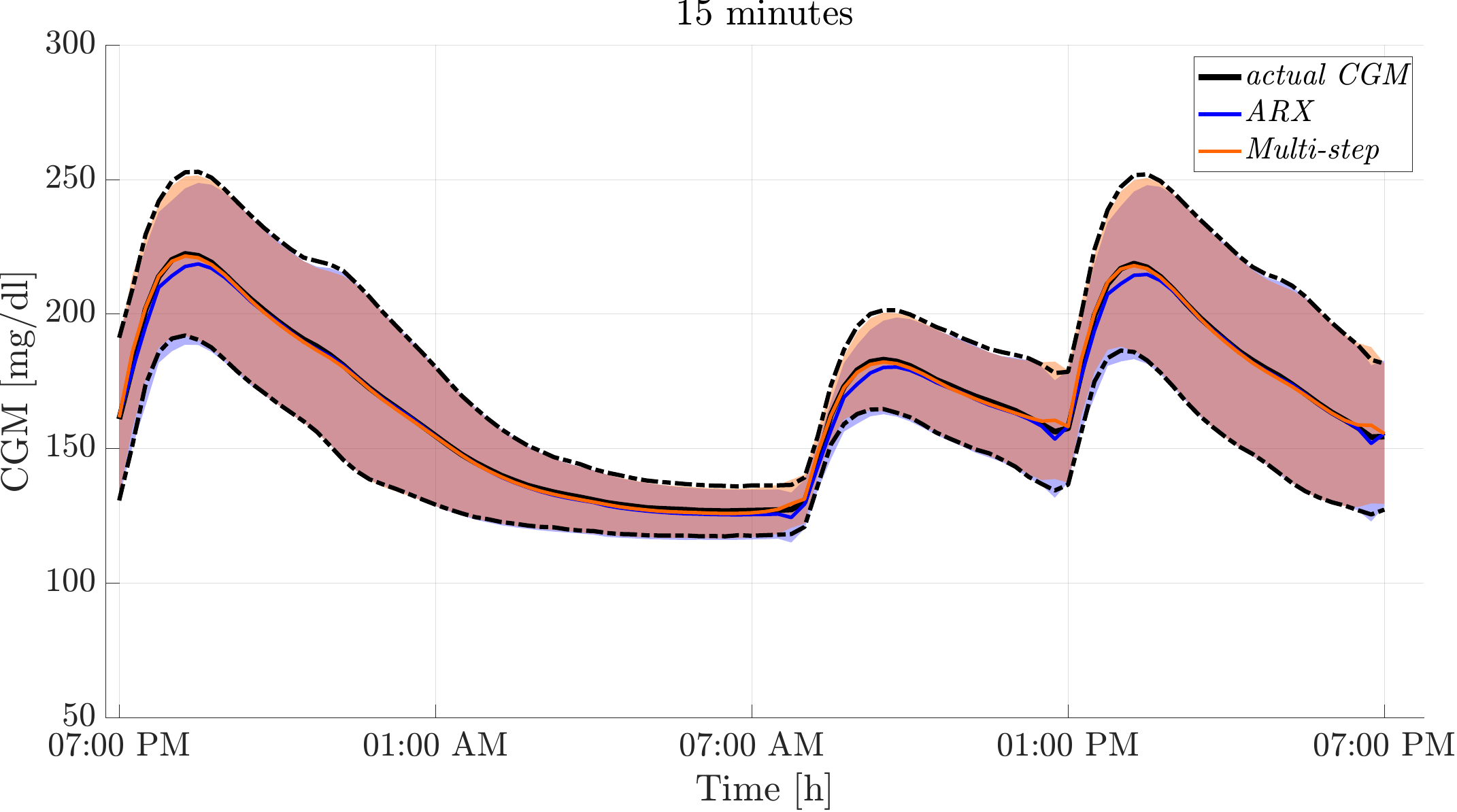}
    \vspace{0.2cm}
    \includegraphics[width=0.48\textwidth]{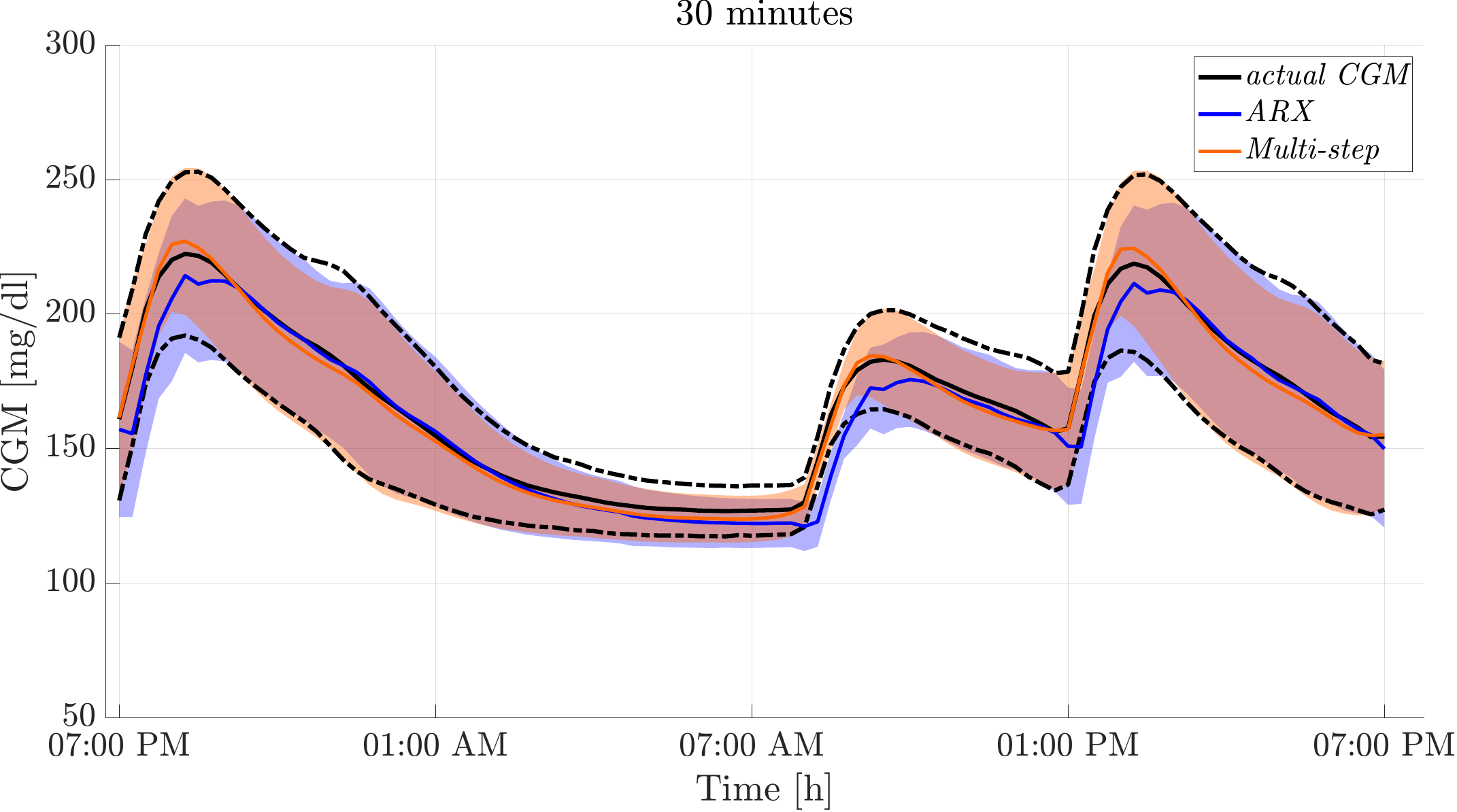}
    \vspace{0.2cm}
    \includegraphics[width=0.48\textwidth]{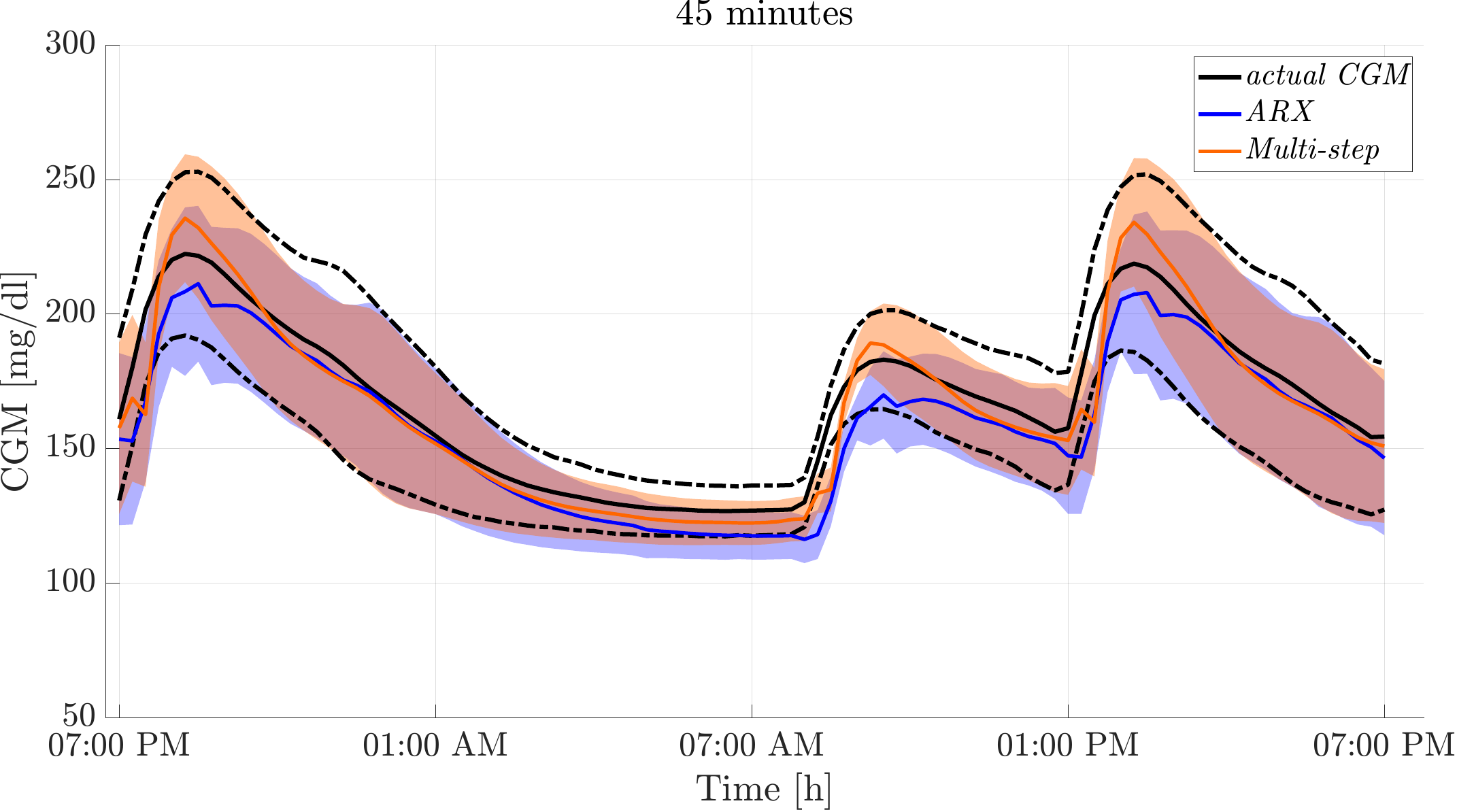}
    \vspace{0.2cm}
    \includegraphics[width=0.48\textwidth]{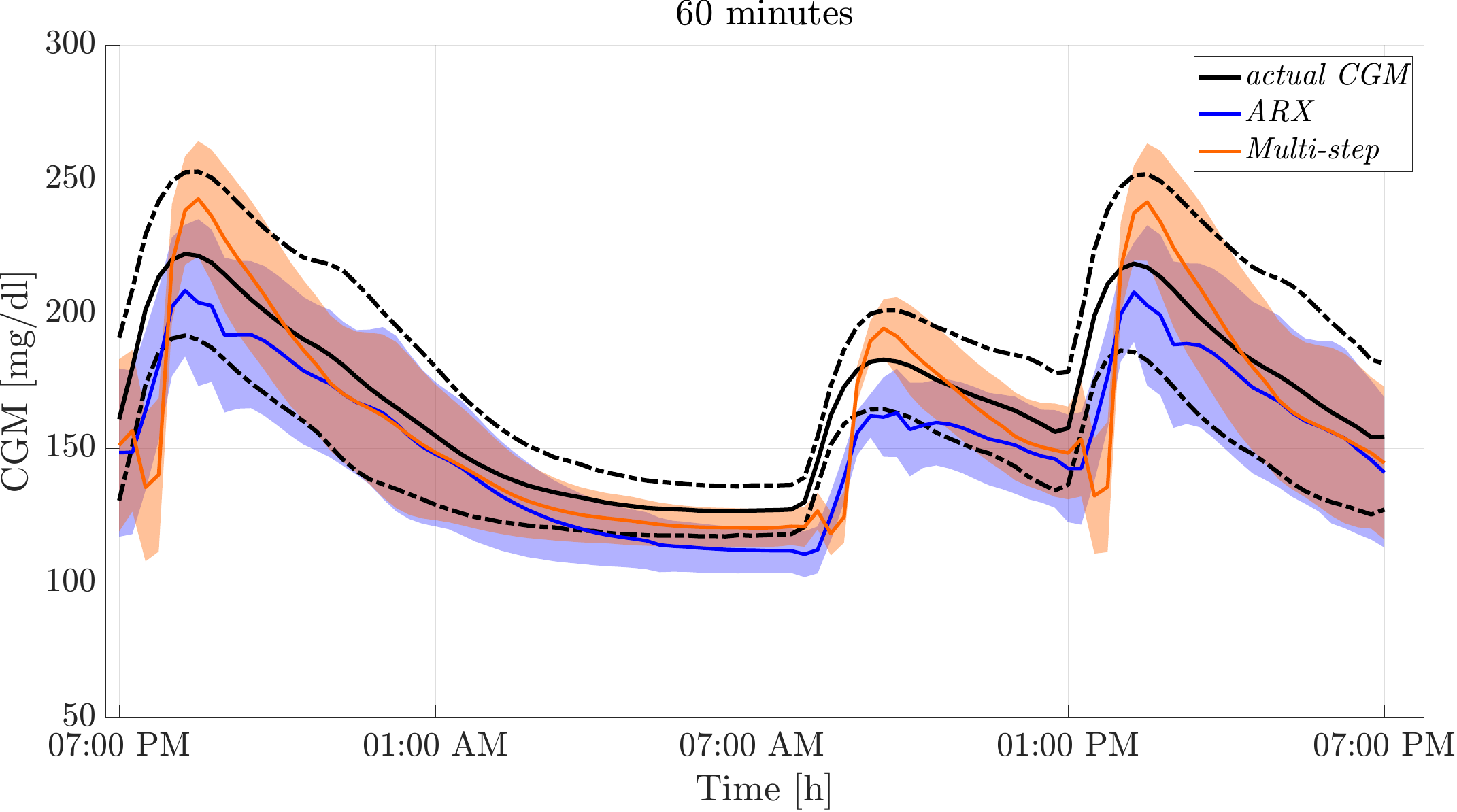}
    \vspace{0.2cm}
    \includegraphics[width=0.48\textwidth]{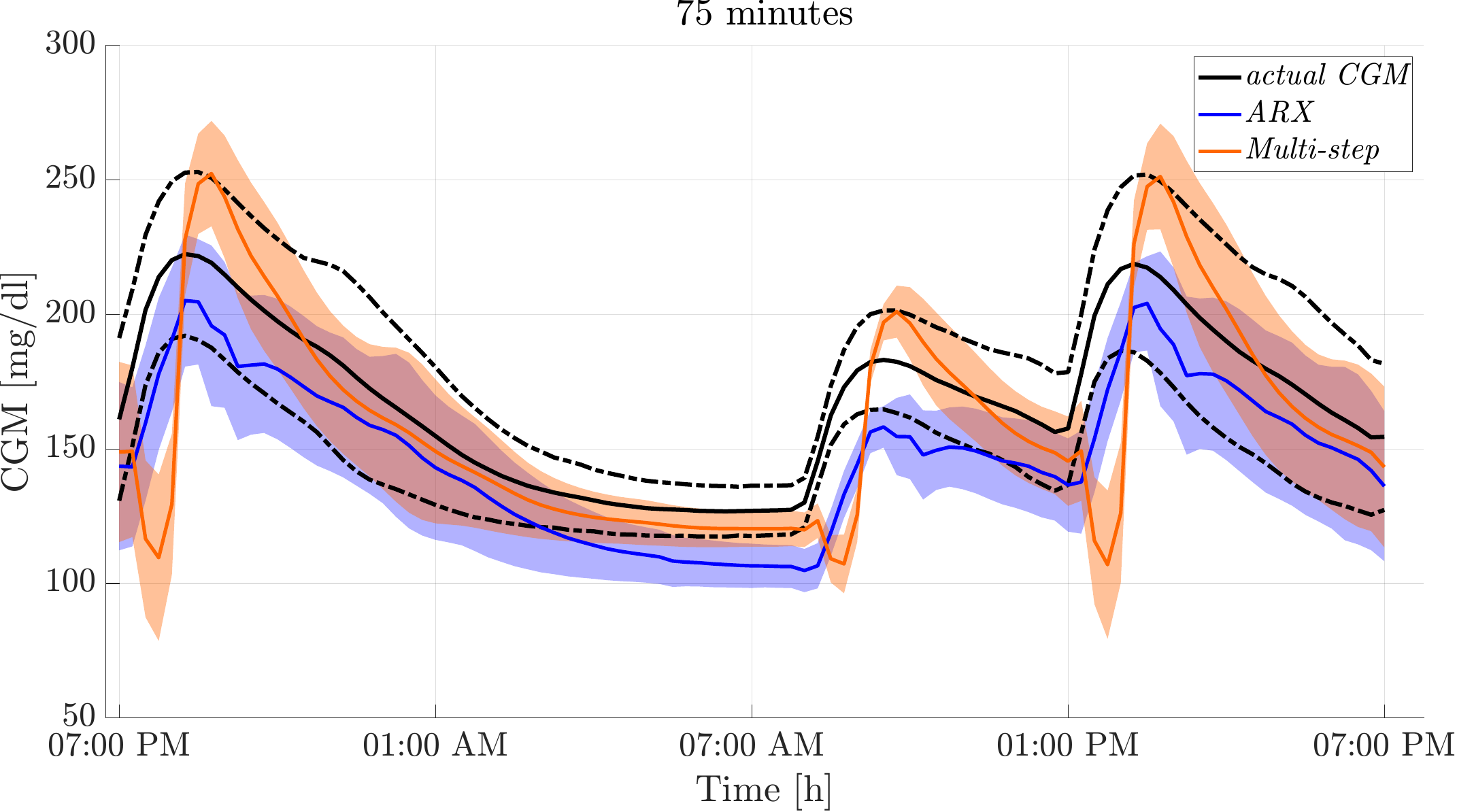}
    \vspace{0.2cm}
    \includegraphics[width=0.48\textwidth]{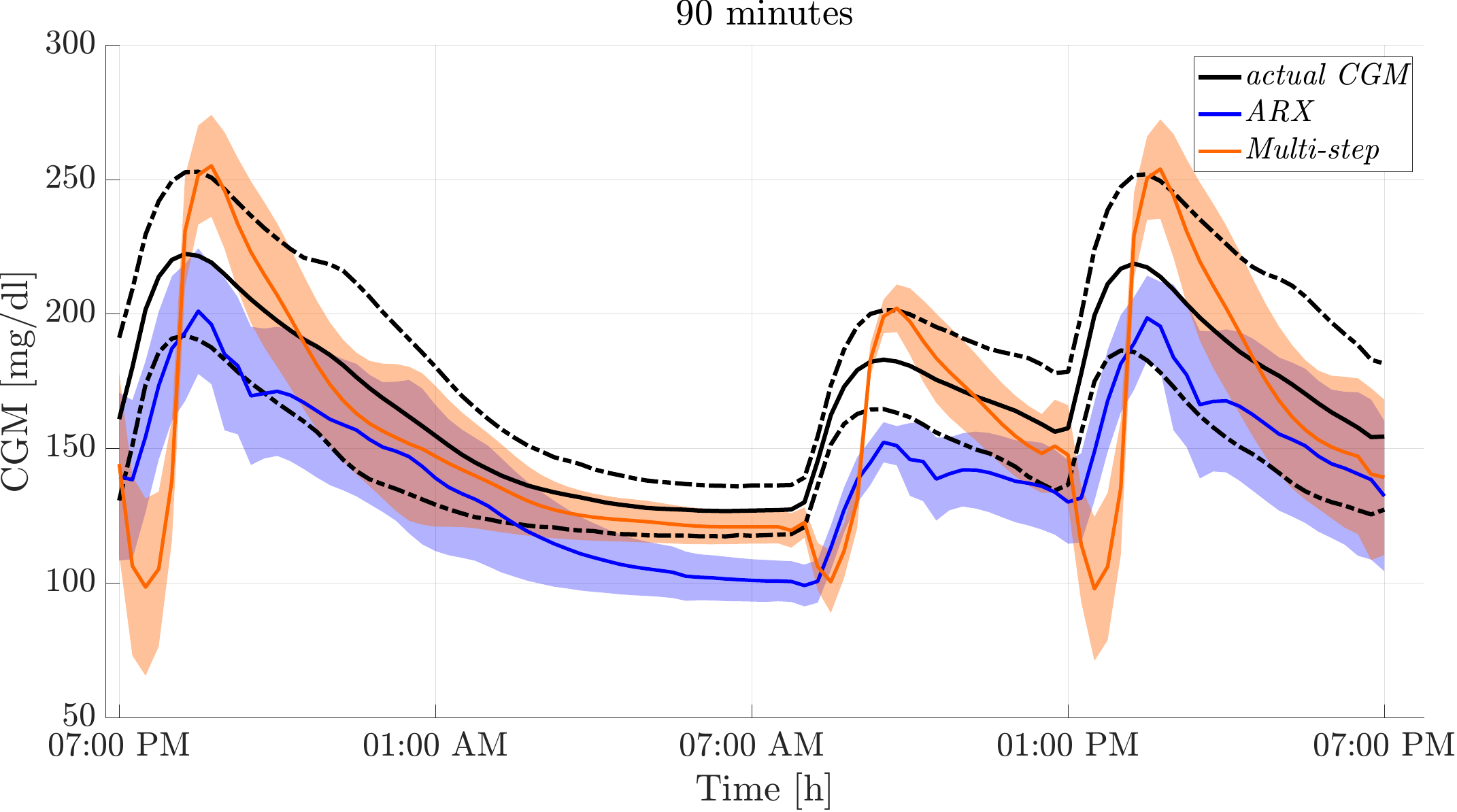}
    \includegraphics[width=0.48\textwidth]{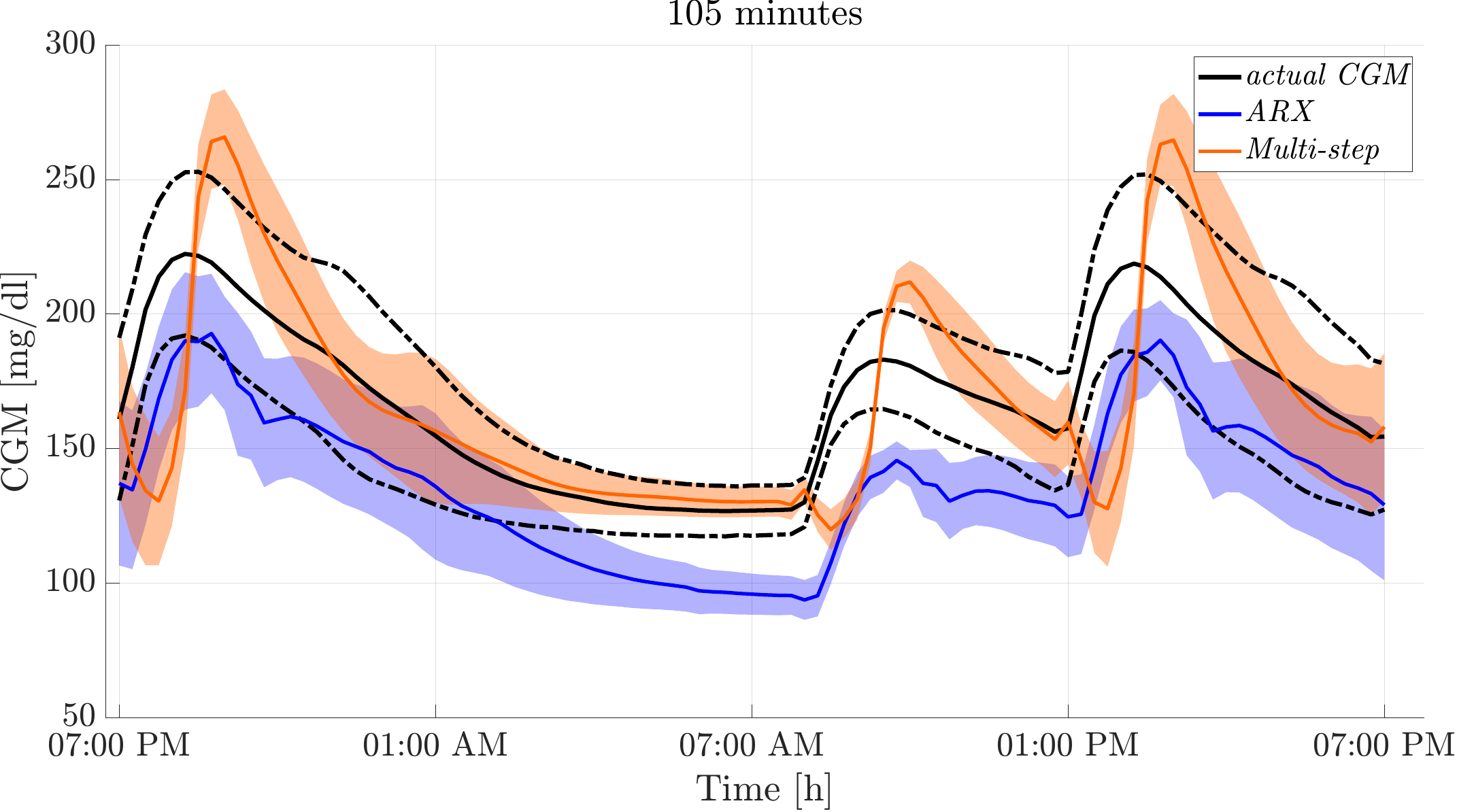}
    \includegraphics[width=0.45\textwidth]{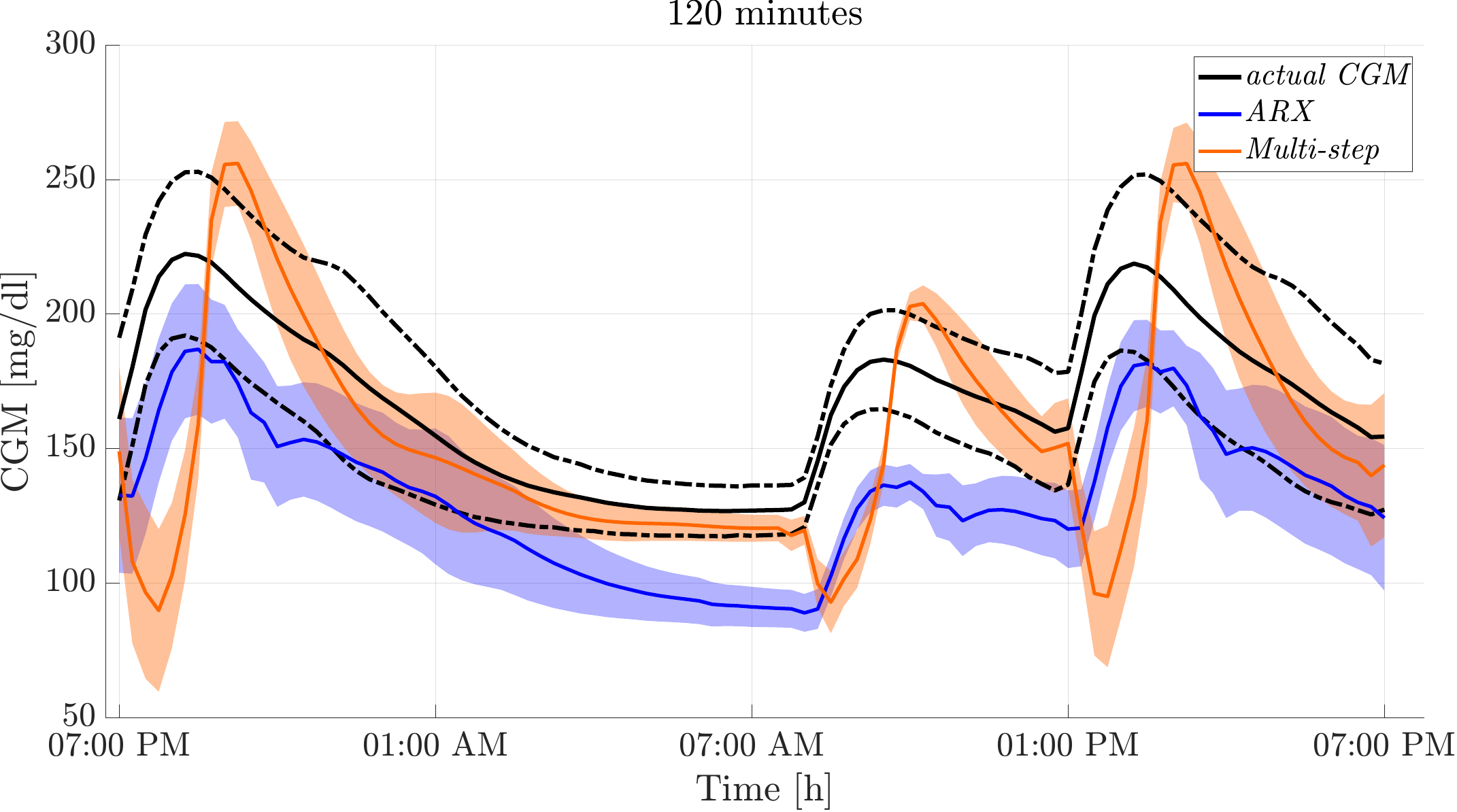}
        \caption{Mean and standard deviation of the predicted blood glucose concentrations obtained by the the 1-step ahead ARX predictor (orange) and the proposed multi-step ahead affine predictors (blue) on the validation dataset. The black solid line represents the mean of the actual CGM measurements, and the black dashed line represents the standard deviation of the actual CGM measurements. Model predictions are reported for each prediction step $T_s \cdot j =15, 30 \dots, 120$ minutes}
    \label{fig:prediction120}
\end{figure*}
\subsection{Identification of the ARX predictor for comparison purposes}
For comparison purposes, we considered a linear, 1-step-ahead predictor because it represents the most widely applied solution in clinical applications for automated insulin delivery systems ~\cite{del2014first,buckingham2018safety,brown2019six,brown2021multicenter,tauschmann2016home,garg2017glucose,deshpande2022feasibility,ozaslan2022feasibility}. 

We identified a linear, 1-step-ahead predictor of blood glucose evolution with an autoregressive model with exogenous inputs describing the system dynamics around a nominal working point.
        Hence, denoting with $\delta u^{ins}=u^{ins}-\overline{u^{ins}}$ the injected insulin variation with respect to the nominal insulin basal rate, $\overline{u^{ins}}$, and $\delta d^{cho}=d^{cho}-\overline{d^{cho}}$ the ingested carbohydrate with $\overline{d^{cho}}=0$, and $\delta y^{cgm}=y^{cgm}-\overline{y^{cgm}}$ the measured glucose variation with the nominal basal glucose, $\overline{y^{cgm}}$, the predicted glucose variation $\hat{\delta y}_k^{ARX}$ is obtained as:
\begin{align*}
    \hat{\delta y}_k^{ARX} = &a_1 \delta y^{cgm}_{k-1} + \dots + a_{n_a} \delta y^{cgm}_{k-n_a}+\\
    &b_{11}\delta d^{cho}_k+\dots+b_{1 n_b}\delta d^{cho}_{k-n_b-1}\\
    &b_{21}\delta u^{ins}_{k-n_k}+\dots+b_{2 n_b}\delta u^{ins}_{k-n_k-n_b-1}
\end{align*}
where  $n_a$ is the number of past outputs, $n_b$ is the number of past inputs, and $n_k$ is the pure time delay. 
The proposed ARX model uses $n_a$:=3, and $n_b$:=3, and $n_k$:=1, respectively, to achieve a flexible, yet parsimonious, description of the system dynamics \cite{toffanin2017mpc, van2011control}.
Using the Prediction Error Method (PEM), from Matlab System Identification Toolbox \cite{ljung1991issues}, the parameters of the proposed $ARX$ are identified, using the identification data from \textit{Scenario-II}, for consistency with the training of the $T$-step-ahead predictor. The identified parameters are:
\begin{equation}
\begin{array}{cccc}
&a_1=-2.20 &   a_2 =1.67 &  a_3=-0.46   \\
&b_{11}=2.09e^{-6}   & b_{12}=7.36e^{-5} &  b_{13}=1.93e^{-4}  \\
&b_{21}=-6.20e^{-5} &   b_{22}=-3.41e^{-3} &  b_{23}=-9.08e^{-4} \\
\end{array}
\end{equation}
The ARX has a sampling time $T_s=15$ minutes, that is the sampling time chosen for the control action, as described in Section 2.1. 
To design a linear MPC, a state space realization of the proposed ARX is required. In particular, we chose the realization corresponding to the canonical controllability form \cite{seborg2016process}. 
The state vector of the ARX model is not measurable, thus a steady-state Kalman filter has been incorporated \cite{rawlings2017model}.

The process noise covariance $Q^{KF}$ is set to the identity matrix, $I_3$, and the measurement noise covariance $R^{KF}$ is defined equal to $1e^{-6}$.  

In the MPC formulation, a standard quadratic cost function was used, with weights $q^{ARX}$ for the system output tracking error $\hat {\delta y}^{ARX} - \overline{\delta y} $ where $\overline{\delta y}$ is the glucose variation with respect to the set points, i.e. $\overline{\delta y}= \bar{y}_T-\overline{y^{cgm}}$. Consistently with the cost function defined in Eq. \ref{eq:J}, the weight $r^{ARX}$ always penalizes the deviation of the control action from the basal rate, i.e. $\delta u^{ins}$. In this formulation, the MPC parameters are defined as $q^{ARX}=1$,  $r^{ARX}=1.5$, and the prediction horizon $T^{ARX}$ is set equal to $T=8$. 
The input and output constraints are consistent with those defined in Eq. \ref{eq:constraints}.

We conduct statistical analyses for each metric and scenario to evaluate the significance of the difference between controller designs. The significance is evaluated based on the average outcomes per subject via paired t-test using two-sided p-value $\le$ 0.05 significance threshold for N = 10. 
\section{RESULTS}\label{sec:results}
\subsection{Accuracy of predictions}
We compare the performance of our proposed multi-step predictor with that of the ARX model-based predictor on the validation dataset (\textit{Scenario-III}) as a function of the step ahead index, i.e., $j=1,\dots,8$.
Fig.~\ref{fig:prediction120} illustrates the predicted population-level CGM time-series on a representive day of the validation dataset. Meals occur at 7 PM, at 8 AM, and 1 PM, respectively.  Generally speaking, an increase in the step ahead index leads to a deterioration in the accuracy of the prediction for a given model, as shown in Figure \ref{fig:prediction120}.
With 1-step ahead, i.e. 15 minutes, the ARX and multi-step predictions are overlapping with the CGM data: this results is expected because the ARX predictor was identified to maximize the 1-step ahead prediction performance. With 2-step ahead, i.e. 30 minutes, the ARX and multi-step predictions are almost overlapping, but the prediction of the ARX model slightly tends to underestimate the glucose excursion around the postprandial phase. This holds also for 3-step ahead case, i.e. 45 minutes, where the ARX model starts to understimate the glucose level also during the nocturnal period. With 4-step ahead, i.e. 60 minutes, the predictions capabilities of both predictors start to degrade. Around meals, the multi-step predictor shows a faster response to insulin bolus than to meal intake. From 75 up 120 minutes, the aforementioned issues become more evident: the ARX predictor heavily underestimates the glucose levels, while the multi-step predictor captures the glucose trend but with an wider excursion around the meals. 
In particular, it is important to note is that the proposed multi-step predictor is able to capture the downward slope of a prandial hyperglycemic excursion as well as the overnight steady-state equilibrium glucose concentration consistently well, regardless of prediction horizon.
The mean and standard deviation outcomes across the 10-adult \textit{in-silico} cohort are presented in Table \ref{tab:prediction_performances}.


\subsection{Evaluating MPC closed-loop performance}
Both MPC were tuned directly using the metabolic model from \cite{kovatchev2020method} with a trial and error approach.
As discussed in Section 5.1, the 1-step ahead ARX predictor tends to underestimate the glucose concentrations, while our multi-step predictor tends to enlarge the dynamic range of glucose responses.  Thus, the best tuning of the ARX-based MPC resulted to be more aggressive with respect to the best tuning of the multi-step-based MPC. We compared the performance of the MPC constructed with the proposed multi-step predictors with those of the ARX-based MPC for each scenario, i.e., \textit{Scenario A}, \textit{Scenario B}, \textit{Scenario C}. In the clinical practice, the effect of the meal disturbance is rejected with a static feed-forward action based on the conventional therapy, which relies on clinical parameters. In our case, to stress the controller capabilities, although the meal is announced to the controller, no additional feed-forward control action is included in the control scheme and the both MPC are completely in charge of disturbance rejection. 

\begin{figure*}[t]
        \centering
        \includegraphics[clip,width=\textwidth]{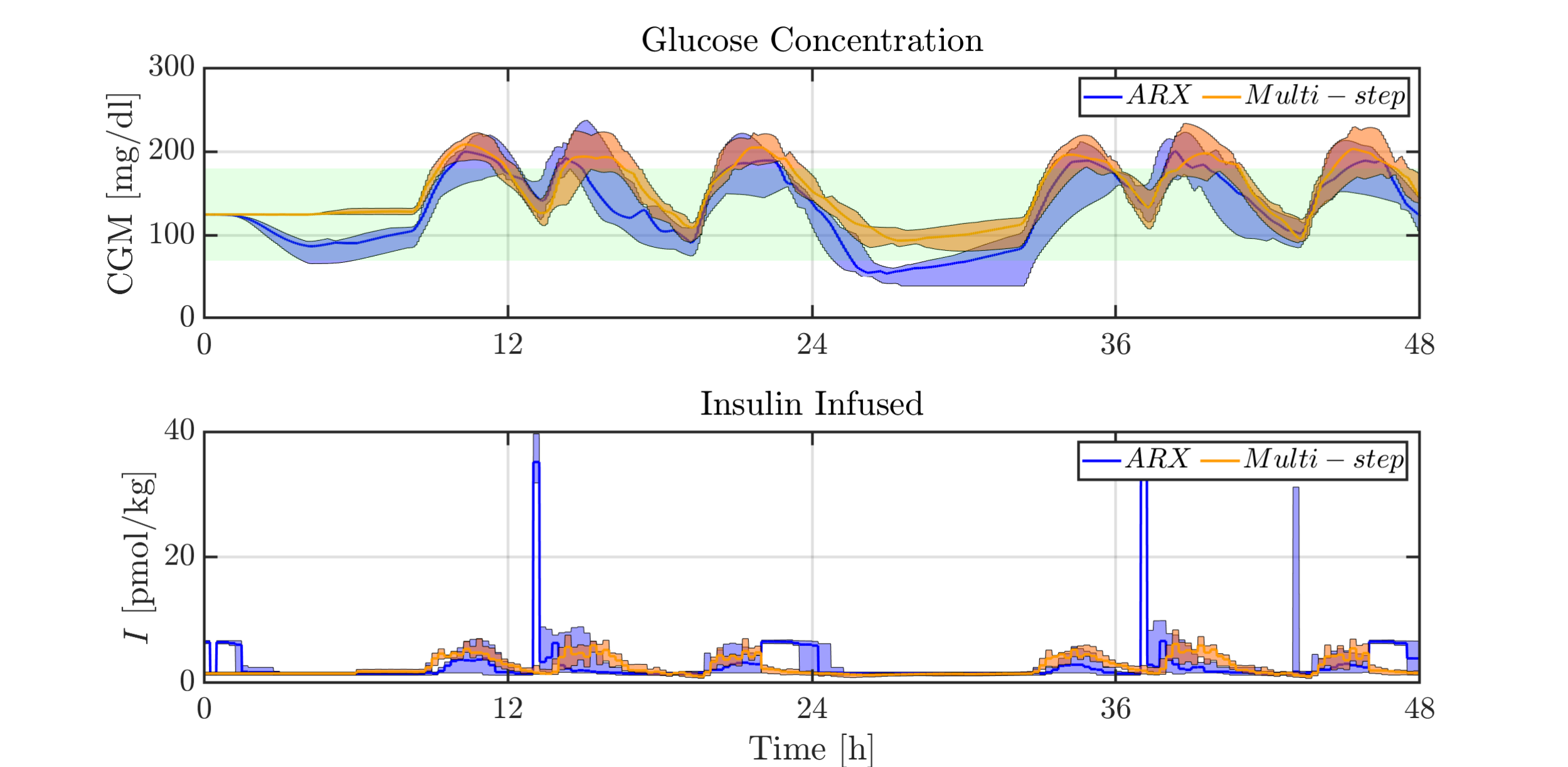 }
        \caption{Scenario A. \textit{Top}: Mean and standard deviation of glucose concentration [md/dL] using the proposed MPC based on the multi-step predictor (yellow) vs. the reference MPC based on the 1-step ahead ARX predictor MPC (blue) ; \textit{Bottom}: Mean and standard deviation  of insulin delivery [pmol/kg] using the proposed MPC based on the multi-step predictor (yellow) vs. the reference MPC based on the 1-step ahead ARX predictor MPC (blue). The green area denotes the euglycemic range, 70-180 mg/dL.}
    \label{fig:scenario1}
\end{figure*}

\begin{figure*}[t]
        \centering
        \includegraphics[clip,width=\textwidth]{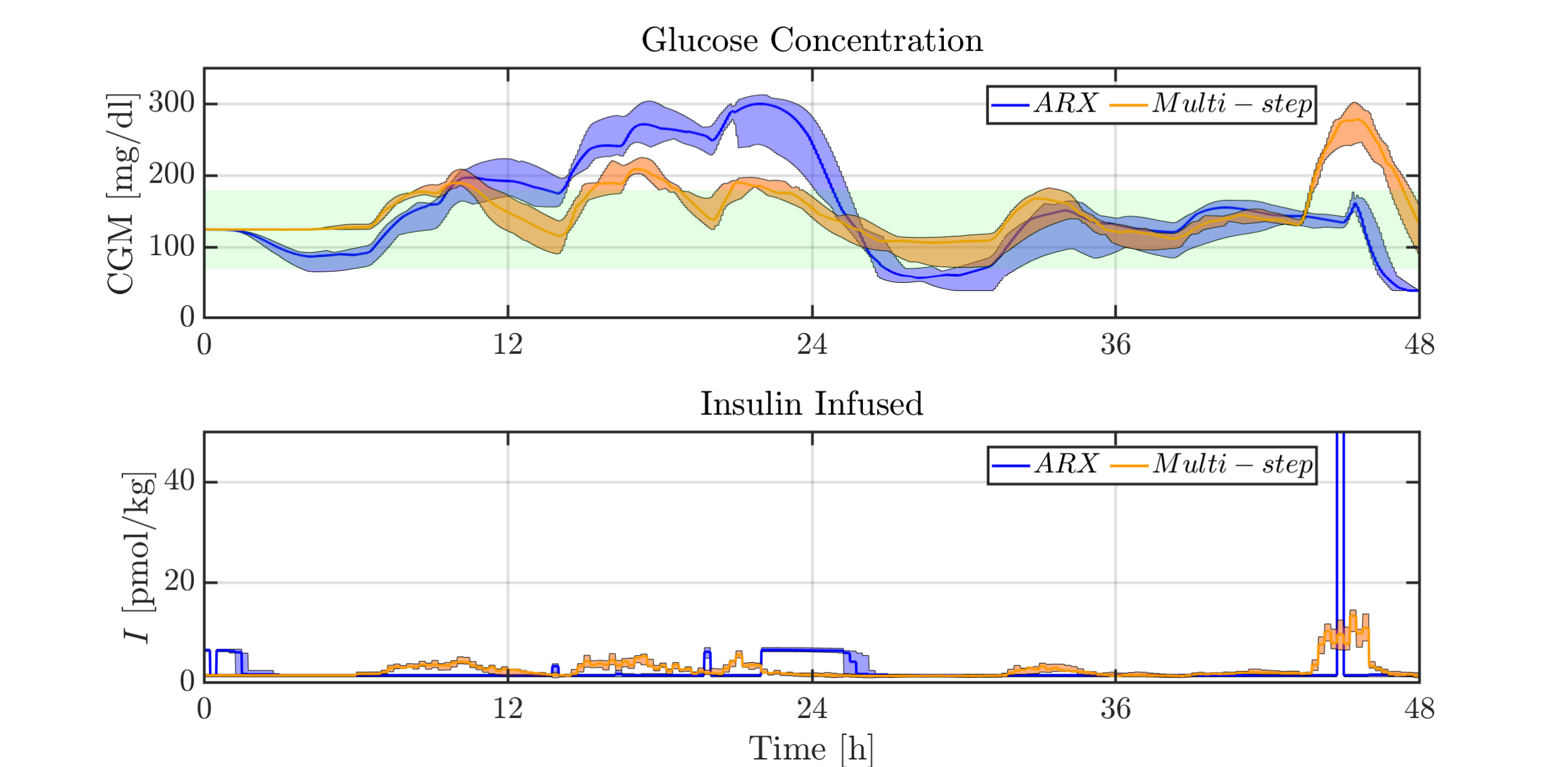}%

        \caption{Scenario B. \textit{Top}: \textit{Top}: Mean and standard deviation of glucose concentration [md/dL] using the proposed MPC based on the multi-step predictor (yellow) vs. the reference MPC based on the 1-step ahead ARX predictor MPC (blue) ; \textit{Bottom}: Mean and and standard deviation  of insulin delivery [pmol/kg] using the proposed MPC based on the multi-step predictor (yellow) vs. the reference MPC based on the 1-step ahead ARX predictor MPC (blue). The green area denotes the euglycemic range, 70-180 mg/dL.}
    \label{fig:scenario2}
\end{figure*}

\begin{figure*}
        \centering
        \includegraphics[clip,width=\textwidth]{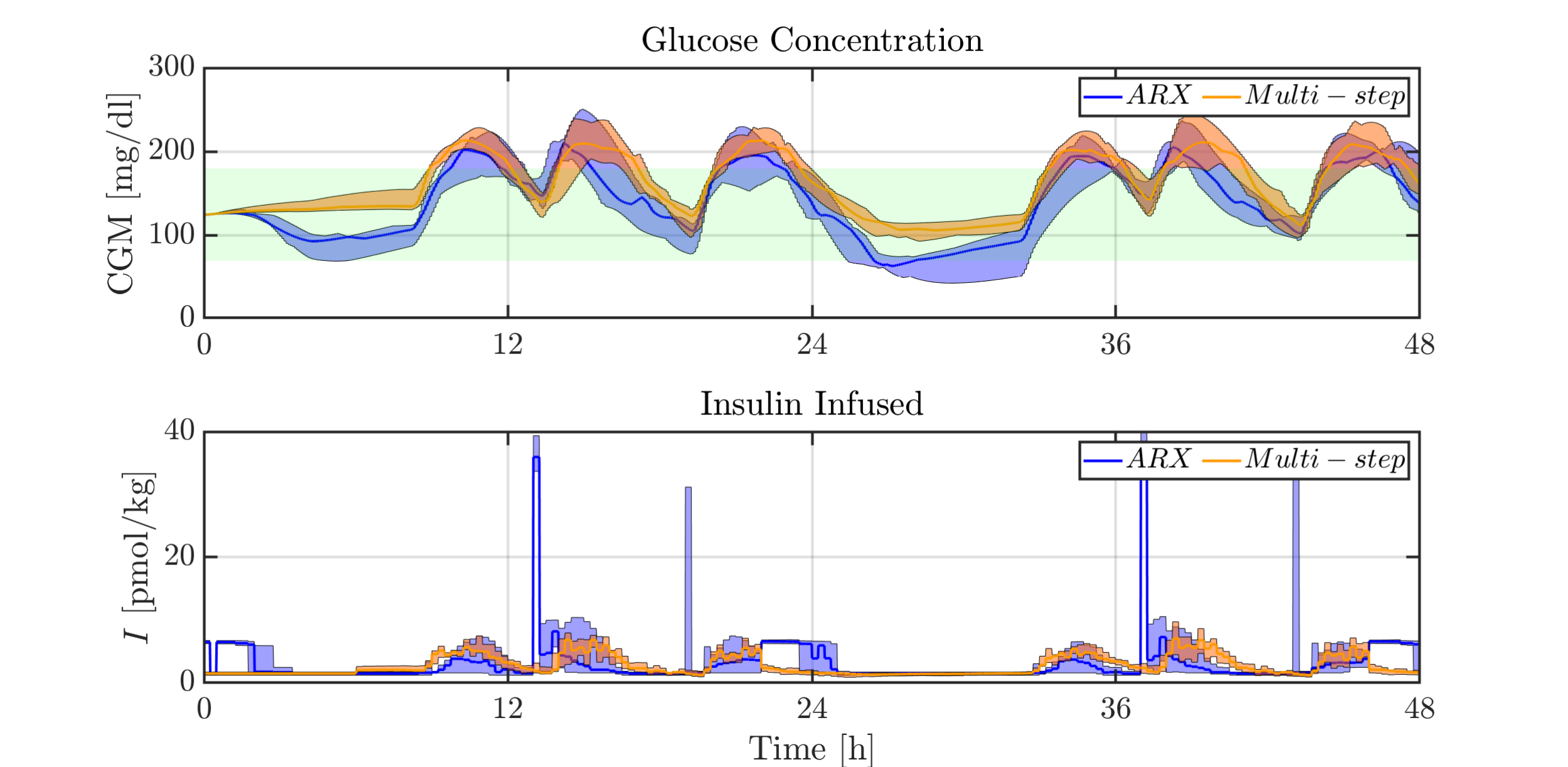}%
        \caption{Scenario C. \textit{Top}: Mean and standard deviation of glucose concentration [md/dL] using the proposed MPC based on the multi-step predictor (yellow) vs. the reference MPC based on the 1-step ahead ARX predictor MPC (blue) ; \textit{Bottom}: Mean and standard deviation of insulin delivery [pmol/kg] using the proposed MPC based on the multi-step predictor (yellow) vs. the reference MPC based on the 1-step ahead ARX predictor MPC (blue). The green area denotes the euglycemic range, 70-180 mg/dL.}
    \label{fig:scenario3}
\end{figure*}

Figures~\ref{fig:scenario1}, \ref{fig:scenario2} and \ref{fig:scenario3} show population-level trajectories on \textit{Scenario A}, \textit{Scenario B}, and \textit{Scenario C}, respectively. 
Compared to the ARX-based MPC, our proposed MPC results in BG curves that are less steep from peak to through, and generally in tighter standard-deviation and min-max envelopes. After correction of meal-induced hyperglycemia, our controller shows a steady insulin delivery with no suspensions. 
In the prandial phase, the ARX-based MPC tends to command extremely large insulin boluses to compensate the meal effect, followed by periods of suspension of insulin delivery, which is an undesirable situation in this application. Suspension of insulin delivery, especially when prolonged, inevitably results in BG rebounds which lead to oscillations and increased glycemic variability and risk of hyperglycemia. It is worthy to note that we could not improve this aspect even with less aggressive tuning of the ARX-based MPC.

Numerical results are tabulated in Table~\ref{tab:performances}, where each block of rows list glycemic metrics for the scenarios, reported as mean and standard deviation across the \textit{in-silico} population. The proposed controller structure yields a \textit{significant} reduction in CV for all simulation scenarios. Further, the percent time spent in hypoglycemia is also \textit{significantly} reduced in all the cases tested. An important observation to make, in this regard, is that the lower hypoglycemia risk is achieved \textit{without} enforcing pump suspensions at all.   
In both \textit{Scenario A} and \textit{Scenario B}, time in severe hyperglycemia is reduced and 
time in the acceptable ranges, i.e., 70-140 mg/dl and 70-180 mg/dl, are increased. However, the time above 180 mg/dl is sligthly elevated, most likely due to a conservative tuning of the controller gains for the prevention of hypoglycemia. In \textit{Scenario C}, the significant reduction in hypoglycemia risk achieved by our proposed controller, is coupled to a slight increase in the time in hyperglycemia and a reduction in the time in acceptable ranges. This tendency to higher glucose level in this case was somewhat expected, since the increase in insulin resistance was unknown to the controller and not modeled for. The notable fact here is that our controller avoided over-delivery of insulin, and prevented the dangerously low glucose levels. 

\begin{table}
  \caption{Glycemic outcomes obtained with the proposed MPC based on the multi-step predictor vs. the reference MPC based on the 1-step ahead ARX predictor. Data is shown as mean (standard deviation) over the \textit{in-silico} population on the three proposed scenarios. P-values were calculated using paired t-test.}
  \label{tab:performances}
  \centering
  \resizebox{0.49\textwidth}{!}{%
  \begin{tabular}{c c| c c c}
   & Metrics & Multi-step & ARX & $p$-value \\
    \hline \hline
     \parbox[t]{2mm}{\multirow{8}{*}{\rotatebox[origin=c]{90}{Scenario A}}}
        & mean [mg/dl] & 151.07 (9.14)  & 129.39 (19.62) & 0.005 \\
        & CV [mg/dl] & 26.22 (5.13) & 40.56 (11.15) & $<$0.001\\
        & $\%$ $<$ 54 [mg/dl] & 0.00(0.00) & 7.24 (10.86)&   0.049\\
        & $\%$ $<$ 70 [mg/dl] & 0.14 (0.37) & 13.98 (12.23) & 0.002  \\ 
        &  $\%$ $\in$ (70,140)  [mg/dl] & 49.32 (7.24) & 45.33(4.02) &  0.145\\
        &  $\%$ $\in$ (70,180)  [mg/dl] & 73.01 (11.02) & 65.31 (7.08) &  0.079\\
        &  $\%$ $>$ 180 [mg/dl]  & 26.84 (11.26)  & 20.69 (9.32) & 0.200 \\ 
        & $\%$ $>$ 250 [mg/dl]  & 0.12 (0.31) & 0.29 (0.50) &  0.381\\ 
\midrule
      \parbox[t]{2mm}{\multirow{8}{*}{\rotatebox[origin=c]{90}{Scenario B}}}
        & mean [mg/dl] & 149.99 (10.47)& 148.47 (19.90) &  0.833 \\
        & CV [mg/dl] & 28.84 (4.94)  & 51.98 (18.99) &  0.001\\
        & $\%$ $<$ 54 [mg/dl] & 1.00 (3.18) & 9.45 (11.71) & 0.409 \\
        & $\%$ $<$ 70 [mg/dl] & 2.26 (4.20)  & 16.20 (14.28) & 0.008 \\ 
        &  $\%$ $\in$ (70,140)  [mg/dl] & 47.78 (8.55)  & 34.62 (9.04)&  0.003\\
        &  $\%$ $\in$ (70,180)  [mg/dl] & 74.99 (7.09)  & 54.15 (14.89) &   $<$0.001\\
        &  $\%$ $>$ 180 [mg/dl]  & 22.74 (7.38)& 29.64 (7.36) & 0.050  \\ 
        & $\%$ $>$ 250 [mg/dl]  & 3.34 (2.46)  & 13.16 (7.78) & 0.001 \\ 
      \midrule
      \parbox[t]{2mm}{\multirow{8}{*}{\rotatebox[origin=c]{90}{Scenario C}}}
& mean [mg/dl] & 161.46 (8.08)& 138.52 (15.68)&  $<$0.001\\
        & CV [mg/dl] & 24.76 (4.84)  & 36.98 (8.33)&  $<$0.001\\
        & $\%$ $<$ 54 [mg/dl] & 0.00(0.00) & 4.89 (7.65) & 0.058 \\
        & $\%$ $<$ 70 [mg/dl] & 0.00(0.00) & 8.71 (9.06)  &0.007  \\ 
        &  $\%$ $\in$ (70,140)  [mg/dl] & 39.19 (8.77)  & 45.70 (5.76) &  0.065\\
        &  $\%$ $\in$ (70,180)  [mg/dl] & 63.88 (8.44) & 66.70 (9.63)&  0.494\\
        &  $\%$ $>$ 180 [mg/dl]  & 36.11 (8.44)  & 24.57 (9.33) & 0.009 \\ 
        & $\%$  $>$ 250 [mg/dl]  & 1.43 (3.11) & 1.15 (1.84) & 0.813 \\ 

\bottomrule
  \end{tabular}
  }
\end{table}
\section{SUMMARY AND CONCLUSIONS}\label{sec:summary}
In this contribution, we proposed a novel closed-loop insulin delivery algorithm for the treatment of T1D. The novelty in our approach lies in the modeling step of the model-based control architecture. We learn a multi-step-ahead predictor of the output which is affine in the control input via a LSTM network. The predictor structure is conceived to increase the accuracy of long-term predictions, allowing at the same time for an efficient formulation and solution of the resulting LTV-MPC. We showed in numerical simulations the better performances of our proposed predictor, compared to that of an ARX-based. As for control performances, we have demonstrated an improvement overall, increased time in target range with significant reduction in glucose variability  and both hypo- and hyperglycemia risks, for a nominal scenario and a scenario testing the robustness against random meal disturbances. In the case of increased insulin resistance, our proposed approach significantly reduced hypoglycemia risk. Planned future work focuses on the design of a more accurate model of the insulin impact on glucose dynamics.


\bibliographystyle{elsarticle-num} 
\bibliography{main}
%
%
%

\end{document}